\documentclass[PRL, reprint,10pt,aps,
, amsmath,amssymb, longbibliography]{revtex4-1}

\usepackage{graphicx}
\usepackage{dcolumn}
\usepackage{bm}

\begin{document}


\title{Hydration lubrication of polyzwitterionic brushes leads to \\ nearly friction- and adhesion-free droplet motion}

\author{Dan Daniel}
 \email{daniel@imre.a-star.edu.sg}
\author{Alfred Yu Ting Chia}
\author{Lionel Chuan Hui Moh}
\author{Rongrong Liu}
\author{Xue Qi Koh}
\author{Xing Zhang}
\author{Nikodem Tomczak}
 \email{tomczakn@imre.a-star.edu.sg}
  
\affiliation{Institute of Materials Research and Engineering, A*STAR (Agency for Science, Technology and Research), 2 Fusionopolis Way, Innovis, Singapore 138634}

\begin{abstract}
Recently, there has been much progress in the design and application of oil-repellent superoleophobic surfaces. Polyzwitterionic brush surfaces are of particular interest, because of their ability to repel oil under water, even in the absence of micro-/nano-structures. The origin of this underwater superoleophobicity is attributed to the presence of a stable water film beneath the oil droplet, but this had not been demonstrated experimentally. Here, using optical interferometric techniques, we show that an oil droplet effectively hydroplanes over a water film, whose thickness is between one hundred and hundreds of nanometres. In addition, using a custom-built Droplet Force Apparatus, we measured the friction and adhesion forces to be in the nN range for millimetric-sized droplets. These forces are much lower than for other classes of well-known liquid-repellent surfaces, including the lotus-leaf effect and lubricant-infused surfaces, where the typical force is on the order of $\mu$N. 
\end{abstract}
\maketitle

\begin{figure}
\includegraphics[]{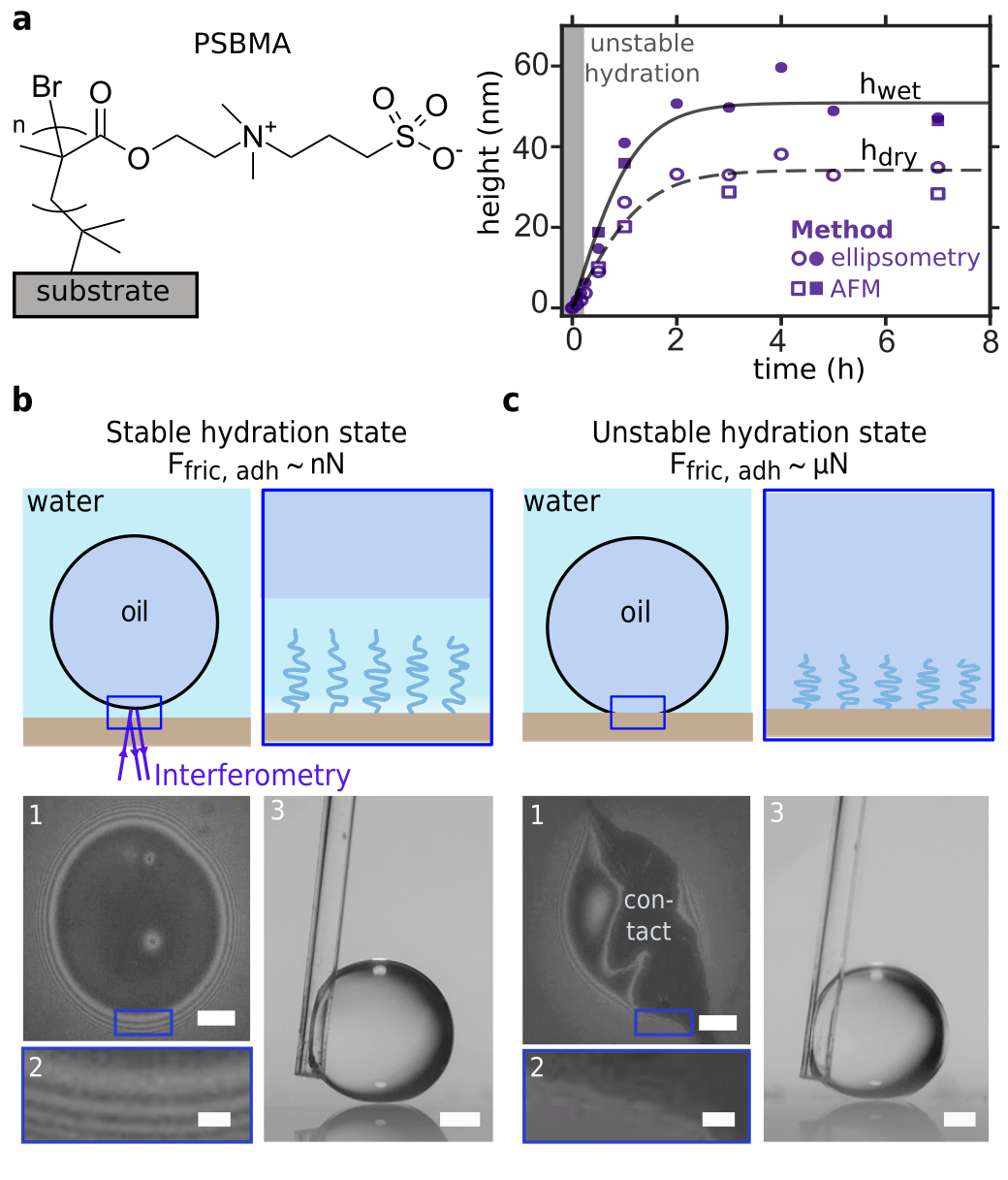} 
\caption{\label{fig:schematic} a) Dry and wet thicknesses of the PSBMA brushes $h_\text{dry, wet}$ measured using ellipsometry and Atomic Force Microscopy (AFM) for different polymerization times. b, c) A stable hydration layer beneath the oil droplet results in ultra-low, nN range adhesion and friction forces $F_\text{adh/fric}$. In contrast, for an unstable hydration layer, $F_\text{adh/fric}$ is on the order of $\mu$N. The presence or absence of a stable water film can be confirmed using Reflection Interference Contrast Microscopy (b-1,2 and c-1,2). In contrast, static contact angle measurements are not sufficient to provide this information (b-3 and c-3). Scale bars are 50 $\mu$m for b-1 and c-1, 10 $\mu$m for b-2 and c-2, and 0.5 mm for b-3 and c-3. The droplets are held in place by a capillary tube.}
\end{figure}

There is a growing interest in developing superoleophobic surfaces for various applications, including oil-repellent coatings and oil-water separation membranes \cite{yong2017superoleophobic}. Most superoleophobic surfaces (both in air and under water) require careful design of micro-/nano-structures to trap an air/water layer and hence confine oil droplet contact to the topmost tips of the structures \cite{tuteja2007designing, Liu1096, liu2009bioinspired}. Damage to the structures greatly impairs the oil repellency performance of these surfaces, as the air/water layer is no longer stable and the oil droplet becomes highly pinned. This is analogous to the wetting transitions from Cassie-Baxter to Wenzel states in lotus-leaf effect superhydrophobic surfaces \cite{de2004capillarity, quere2008wetting}. 

In contrast, polyzwitterionic brush surfaces are able to repel oil under water even in the absence of micro-/nano-structures \cite{kobayashi2012wettability, liu2017nature}. This has been attributed to the ability of polyzwitterionic brushes to swell under water and retain hydration shells around their charged moieties. This stabilizes a water film between the oil droplet and the underlying substrate, and hence results in extreme oil-repellency. Incidentally, it is thought that the presence of a water film on cartilage surfaces (stabilized by zwitterionic phospholipids) is similarly responsible for the ultra-low friction in skeletal joints \cite{jahn2016lubrication, zhulina2014lubrication}. 

Despite recent progress, there is little direct experimental evidence for the presence of the water film between the oil droplet and the substrate, and most previous discussions implicitly assume the presence of such a hydration layer. \cite{kobayashi2012wettability, liu2017nature, huang2015surface, shi2016long}. The thickness of this water film---should it exist---is not known and its relation to oil-repellency remains poorly understood. Furthermore, the effect of such a water film on adhesion and friction of oil droplets has not been investigated. Here, using Reflection Interference Contrast Micrscopy (RICM), we are able to directly visualize and quantify the thickness of the water film between an oil droplet and a glass substrate grafted with poly(sulfobetaine methacrylate) (PSBMA) brushes. In addition, using a custom-built instrumentation, which we name the Droplet Force Apparatus (DFA), we found the friction and adhesion forces of the oil droplet $F_\text{adh/fric}$ to be in the nN range. Our results can be largely explained by viscous dissipation in the experimentally observed water film. Finally, we propose and experimentally verify simple scaling relations for $F_\text{adh/fric}$. 

\section*{Results}
The PSBMA brushes were grown on glass or silicon wafer using Surface Initiated Atom Transfer Radical Polymerization (SI-ATRP) \cite{azzaroni2006ucst, zoppe2017surface}. The brush thickness, as measured with ellipsometry and Atomic Force Microscopy (AFM), initially increased linearly as a function of the polymerization time, but then reached a plateau after about an hour to a maximum thickness $h_\text{dry} = 34 \pm 4$ nm (Fig.~\ref{fig:schematic}a). When immersed under water, the brush swells by about 50$\%$ to $h_\text{wet} = 51 \pm 5$ nm. The relatively small swelling ratio is consistent with densely grafted chains in the brush state \cite{azzaroni2006ucst, cheng2008thickness}. See Supplementary Figures S1--S3 for contact angle measurements, chemical, and AFM characterizations of the brushes.   

The presence or absence of a water film can be confirmed using Reflection Interference Contrast Microscopy (RICM) \cite{limozin2009quantitative, daniel2017oleoplaning}. Here, we shone monochromatic light (wavelength $\lambda$ = 561 nm) from below and captured the reflection off the droplet's base using a camera (Fig.~\ref{fig:schematic}b, c). The presence of a water film results in thin-film interference and, in particular, dark and bright fringes as light reflected off the various interfaces (oil-water, water-brush, and brush-glass) interfere destructively and constructively with one another.

\begin{figure*}
\includegraphics[width=\textwidth]{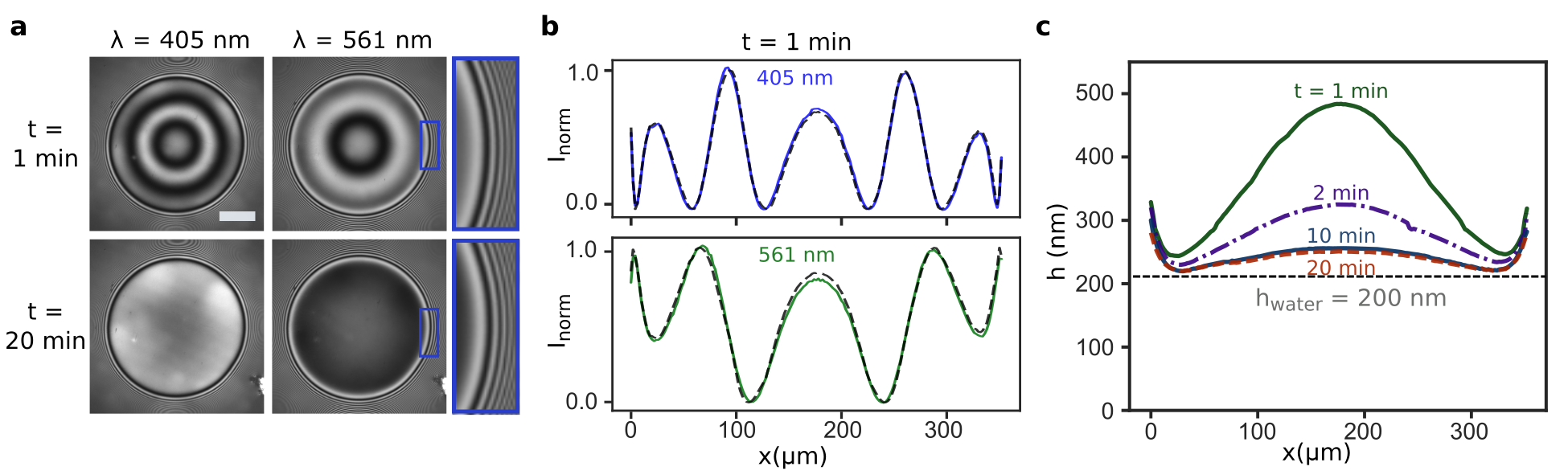}
\caption{\label{fig:RICM} a) Water film under droplet visualized using confocal RICM at wavelengths $\lambda =$ 405 and 561 nm at different times. Scale bar is 100 $\mu$m. b) Normalized reflection light intensities across the centre of the droplet's base at $t = 1$ min for $\lambda =$ 405 and 561 nm (solid blue and green lines, respectively). Dashed lines are the best-fit lines using theory of Fresnel reflection. c) Water film profiles at different times. The equilibrium film thickness $h_{\text{water}}$ is about 200 nm.
 }
\end{figure*}

For a glass surface grafted with 30-nm-thick PSBMA brushes, the presence of interference fringes at the edge of the droplet's base (Fig.~\ref{fig:schematic}b-1,2; see also Figure \ref{fig:RICM}a) indicates the presence of a \textit{continuous} water film. Since there is no contact between the droplet and the underlying substrate, the droplet's base is free from pinning and hence appears smooth and circular (Fig.~\ref{fig:schematic}b-1). In contrast, for 6-nm-thick brushes, the water film is not stable and the droplet-solid contact area can clearly be seen in Figure \ref{fig:schematic}c-1; in place of interference fringes at the edge, there is instead an irregularly shaped outline due to three-phase contact line pinning (Fig.~\ref{fig:schematic}c-2).

We note that the static contact angles of the oil droplet can be indistinguishable (effectively 180$^{\circ}$) for both stable and unstable states (Fig.~\ref{fig:schematic}b-3 and c-3). Recently, several groups have established the difficulty of accurately measuring contact angles close to 180$^{\circ}$ \cite{srinivasan2011assessing, Liu1147}. Small uncertainty in the baseline of the droplet, even at single pixel level, leads to large error in contact angle values ($>10^{\circ}$), highlighting the challenge in using conventional contact angle measurements to characterize the wetting properties of surfaces \cite{srinivasan2011assessing, Liu1147, daniel2018origins, huhtamaki2018surface}

Experimentally, we found that the water film becomes unstable for brushes with thicknesses $h_{\text{dry}} < 5$ nm and $h_{\text{wet}} < 9$ nm (shaded gray, Fig.~\ref{fig:schematic}a). This is likely because for short polymer chains, the wetting properties of the surface are dominated by the presence of non-hydrophilic bromine atoms from the polymerization initiator. X-ray photoelectron spectroscopy (XPS) characterization of the surface is consistent with this view (Supplementary Fig. S2).

The presence of a stable water film has a profound impact on the oil-repellent performance of the PSBMA brushes. The measured adhesion and friction forces $F_{\text{adh/fric}}$ for a millimetric-sized oil droplet are on the order of tens of nN for a stable hydration state, but increase to several $\mu$N for an unstable hydration state. In comparison, $F_{\text{adh/fric}}$ is on the order of $\mu$N for similarly-sized water droplets on  superhydrophobic surfaces exhibiting the lotus-leaf effect \cite{liimatainen2017mapping, daniel2018origins, reyssat2009contact}.

\begin{figure*}
\includegraphics[width=\textwidth]{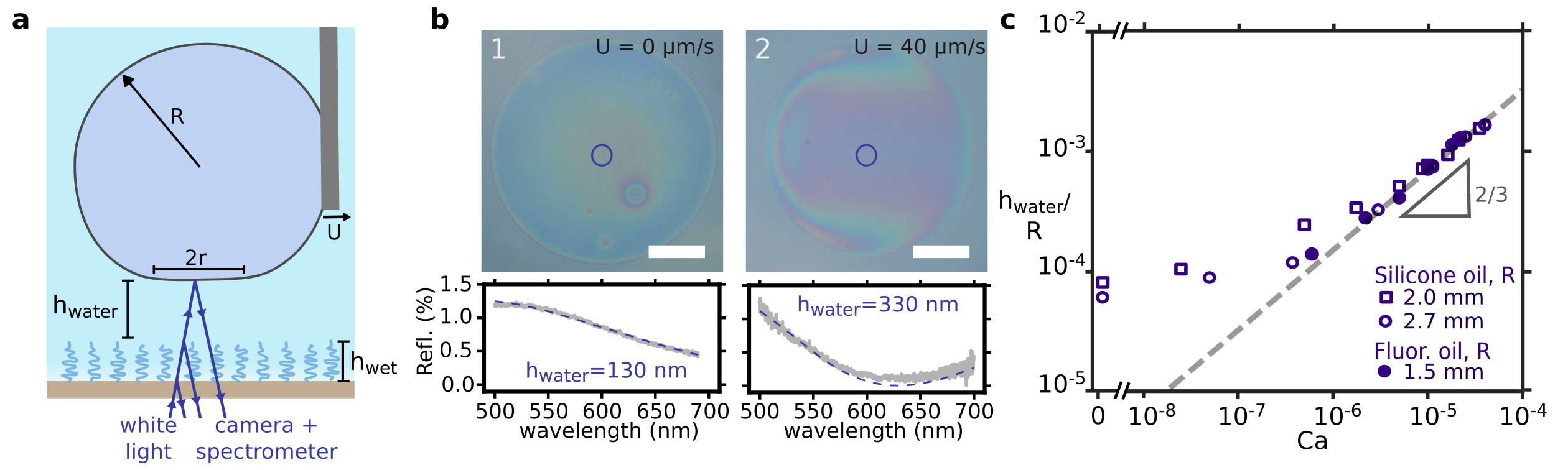} 
\caption{\label{fig:thickness} a) Water film thickness $h_{\text{water}}$ measured using spectroscopic reflectometry (See Supplementary Figures S4 and S5 for details of the setup). b) Thin-film interference due to the water film results in different colours observed under the droplet. Absolute values of $h_{\text{water}}$ can be determined with nanometric resolutions by fitting the reflectivity measurements from the spectrometer (measured areas are indicated by blue circles) with classical theory of Fresnel reflection \cite{born2013principles}. Dashed-lines are the best-fit curves. Scale bar is 100 $\mu$m. c) Non-dimensionalized quantities $h_{\text{water}}/R$ against $Ca = \eta U/\gamma$ obtained for silicone and fluorinated oils. Dashed line is the prediction from LLD (Equation \ref{eq:LLD}), with a prefactor of 1.5.  
 }
\end{figure*}

For a stationary droplet, the water film thickness $h_{\text{water}}$ can be determined using dual-wavelength confocal RICM \cite{daniel2017oleoplaning, limozin2009quantitative}. We raster scanned the surface simultaneously with two focused beams of monochromatic lights with wavelengths $\lambda$ = 458 and 561 nm, and captured the reflected light through the pinhole of a confocal microscope (Fig.~\ref{fig:RICM}a). From the reflection intensities of the two wavelengths, the water film profile $h_{\text{water}}$ can be deduced unambiguously using classical Fresnel theory (Fig.~\ref{fig:RICM}b). After waiting for 20 minutes, we find that the equilibrium $h_{\text{water}}$ is about 200 nm, much thicker than the swollen brush film thickness $h_{\text{wet}}$ (Fig.~\ref{fig:RICM}c).

Alternatively, $h_{\text{water}}$ can be measured by shining white light and analyzing the reflected light using a spectrometer, i.e. spectroscopic reflectometry. Similar equilibrium film thicknesses (between 100--200 nm) are obtained using this method. Details of the two techniques (spectroscopic reflectometry and RICM) can be found in our previous work \cite{daniel2017oleoplaning} and also in Supplementary Figures S4--S8. See also discussion in Materials and Methods.    
 
The relatively thick water film is stabilized by repulsive electric double-layer force. In water, both oil and the PSBMA surface acquire negative surface charges, with reported zeta potentials of -40 mV and -35 mV, respectively \cite{shi2016long, guo2015surface}. In the absence of added electrolytes, the electrostatic forces between the charged interfaces generate a repulsive pressure given by
\begin{equation} \label{eq:P}
\begin{split}
P(h) = 2 \epsilon_{o} \epsilon (\pi k_{\text{B}} T/ze)^2/ h^2,
\end{split}
\end{equation}   
where $\epsilon_{o}$ and $\epsilon = 80$ are the vacuum and relative permittivities, $k_{\text{B}}$ is the Boltzmann constant, $T$ is the absolute temperature, $z$ is the valency of the counterions, and $e$ is the elementary charge \cite{israelachvili2011intermolecular}. $P(h)$ is balanced by the capillary pressure $\gamma/R$, which predicts an equilibrium $h$ of about 100 nm, consistent with the experimentally observed $h_{\text{water}}$. We also note that with added salt (0.1 M NaCl), the dissolved ions will screen the surface charges and we observed that $h_{\text{water}}$ was reduced to 10 nm (See Supplementary Fig.~S8), as predicted by electric double-layer forces (Supplementary Eq.~S2). 

Experimentally, we found that $h_{\text{water}}$ increases with droplet's speed $U$. For example, $h_{\text{water}}$ for the same droplet increases from 130 to 330 nm when moving at $U$ = 40 $\mu$m/s (Fig.~\ref{fig:thickness}b-1,2), and reaches more than 1 $\mu$m for $U > 1$ mm/s (Supplementary Fig.~S5d). For silicone and fluorinated oils moving at $U =$ 0.1--2.0 mm/s, $h_{\text{water}}$ is well described by the Landau-Levich-Derjaguin (LLD) scaling
\begin{equation} \label{eq:LLD}
\begin{split}
h_{\text{water}}/R \sim Ca^{2/3},
\end{split}
\end{equation} 
where $R$ is the droplet's radius, $Ca = \eta U/\gamma$ is the capillary number, $\eta$ is the water viscosity, and $\gamma$ = 40--50 mN/m is the oil-water interfacial tensions (Fig.~\ref{fig:thickness}c) \cite{landau1942, derjaguin1943, bretherton1961motion, daniel2017oleoplaning}. This is analogous to liquid films entrained during the dip-coating process. 


It is known that water remains fluid even when confined to subnanometre films \cite{raviv2004fluidity}. Thus, for $h_{\text{water}} > 100$ nm, water acts as a lubricating film, and we expect friction and adhesion forces $F_{\text{fric, adh}}$ to be minimal and be dominated by viscous dissipation in the water film. Moreover, $F_{\text{fric, adh}}$ should be relatively insensitive to details of the brush layer, such as its thickness $h_{\text{wet}}$ and grafting density. We show this explicitly by measuring $F_{\text{fric, adh}}$ using a custom-built force sensor, which we name the Droplet Force Apparatus (DFA). 

\begin{figure*}
\includegraphics[width=\textwidth]{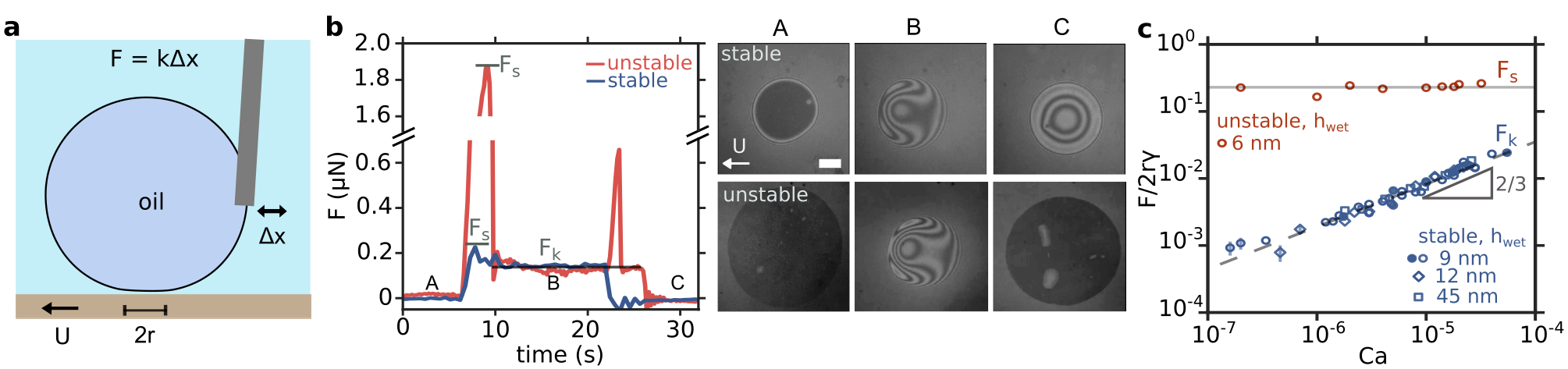} 
\caption{\label{fig:friction} a) Friction force measurement using the Droplet Force Apparatus (DFA). b) Force-time curves for droplets ($2R$ = 2.3 mm) for stable and unstable water films (blue and red lines, respectively). The surfaces started moving with $U$ = 0.7 mm/s at time $t = 8$ s and stopped at $t = 25$ s. Oscillation of $F$ when the motor was stopped are transients due to sudden deceleration akin to a dampled harmonic oscillator. RICM images are for the droplets' base at time points A, B, and C.  Scale bar is 50 $\mu$m. See Supplementary Movies S1 and S2. c) Plots of non-dimensionalized force $F/2r\gamma$ vs. capillary number $Ca$. Each point represents a single force measurement. The uncertainties in deflection detected by the auto-correlation algorithm are either smaller than the point size or represented by the error bars. See Supplementary Fig.~S7 for further discussion.}
\end{figure*}

\begin{figure*}
\includegraphics[width=\textwidth]{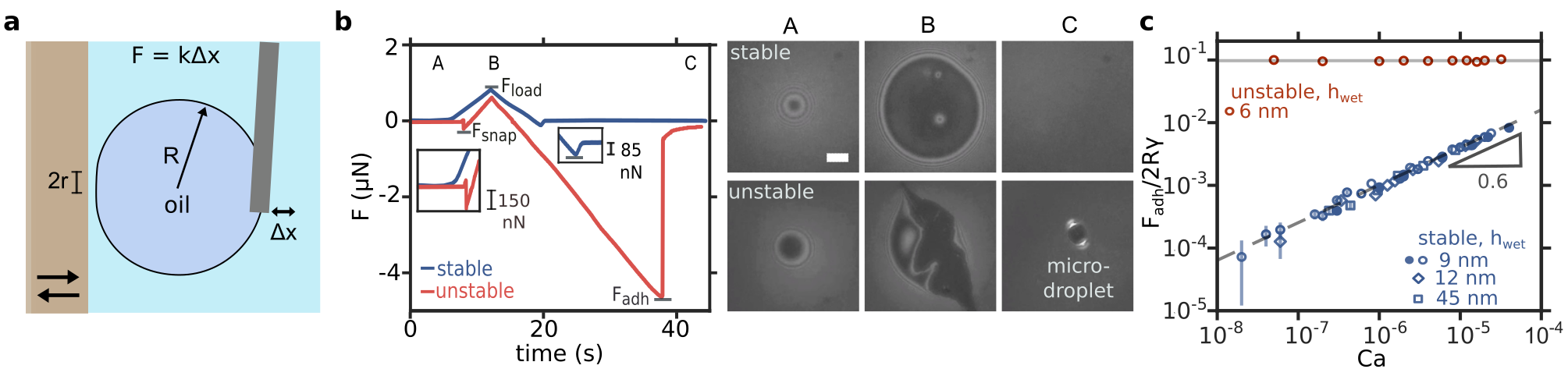} 
\caption{\label{fig:adh} The adhesion force was measured using DFA. b) Force-time curves for droplets ($2R$ = 2.0 mm) for stable and unstable water films (blue and red lines, respectively). The surface was approaching the droplet (A to B) and retracting from it (B to C) at a speed $U$ = 0.05 mm/s . RICM images are for the droplets' base at time points A, B, and C. Scale bar is 50 $\mu$m. See Supplementary Movies S3 and S4. c) Plots of non-dimensionalized force per unit length $F_{\text{adh}}/2R\gamma$ vs. capillary number $Ca$. Each point represents single force measurement. The uncertainties in deflection detected by the auto-correlation algorithm are either smaller than the point size or represented by the error bars. See Supplementary Fig.~S7 for further discussion.}
\end{figure*}

DFA is an improvement to an experimental setup reported in our previous works \cite{daniel2017oleoplaning, daniel2018origins}. By building the whole setup on an active-damped optical table and inside an enclosure to minimize draught, we are able to greatly reduce the environmental noise and measure forces as small as tens of nN, making it one of the most sensitive force sensors for wetting applications \cite{liimatainen2017mapping, pilat2012dynamic}. In contrast, our previous setup has a sensitivity of more than 100 nN \cite{daniel2017oleoplaning}. At the same time, the film dynamics at the droplet's base can be visualized by either RICM (Fig.~\ref{fig:schematic}b) or spectroscopic reflectometry (Fig.~\ref{fig:thickness}) making the combined setup an extremely powerful scientific apparatus for the direct and simultaneous measurement of contact areas, thicknesses, and forces for liquid droplets interacting with various interfaces. Details of the setup can be found in Supplementary Figures S4--S8.  

Figure \ref{fig:friction}a is a schematic of how $F_{\text{fric}}$ was measured using the DFA for different speeds $U$. The droplet was attached to an acrylic capillary tube with inner and outer diameters of 0.288 and 0.360 mm, respectively; while the surface was moving at controlled $U$, the force acting on the droplet was inferred from the tube's deflection $\Delta x$, since $F = k \Delta x$, where $k$ = 2--20 mN/m is the flexular spring constant which we determine individually for each tube. With a high-resolution camera and auto-correlation algorithm, it is possible to detect $\Delta x$ of a few $\mu$m, which translates to a force resolution of about 10 nN (Supplementary Fig.~S6, S7).

Figure \ref{fig:friction}b shows the force-time curves for silicone oil droplets (2$R$ = 2.3 mm) moving at $U = 0.7$ mm/s. When a stable water film is present, the force required to jumpstart the motion $F_{\text{s}} = 230$ nN is close to that required to maintain the motion $F_{\text{k}} = 150$ nN (Supplementary Movie S1). This is reminiscent of the static and kinetic friction forces between two solid surfaces \cite{gao2018drops, daniel2018origins}. For a droplet hydroplaning on a water film that follows the LLD scaling, it is known that the friction force due to viscous dissipation is given by
\begin{equation} \label{eq:F_LLD}
\begin{split}
F_{\text{fric}} \sim 2r \gamma Ca^{2/3},
\end{split}
\end{equation}
where $2r$ is the size of the droplet's base depending on the applied normal force $F_{\text{N}} = (2\gamma/R) \, \pi r^{2}$ \cite{daniel2017oleoplaning, keiser2017drop} (See Supplementary Figure S9 for derivation). For silicone and fluorinated oils (empty and filled markers in Figure \ref{fig:friction}c, respectively) of $2R =$ 3.3--4.7 mm and $2r  =$ 0.25--0.93 mm, as long as the water film is stable, $F_{\text{k}}$ is well described by Equation \ref{eq:F_LLD} (dashed line in Figure \ref{fig:friction}c with a prefactor of 12) and is independent of $h_{\text{wet}}$ = 9--45 nm. 

We can also define the effective coefficient of friction $\mu_{f} = F_{\text{fric}}/F_{\text{N}} \sim (R/r) \, Ca^{2/3}$. For millimetric droplets moving at $U$ = 1--1000 $\mu$m/s, this translates to $\mu_{\text{f}}$ = 0.01--0.1. In comparison, $\mu_{\text{f}}$ can be as low as 0.001 for \textit{solid-solid} friction between cartilage surfaces \cite{jahn2016lubrication, zhulina2014lubrication}. Nonetheless, $F_{\text{fric}}$ is remarkably low. At the lowest speed $U = 10$ $\mu$m/s, $F_{\text{fric, min}}$ = 25 nN, much lower than that for lotus-leaf effect surfaces (typically $\mu$N for speeds ranging from $\mu$m/s to mm/s) \cite{daniel2018origins, liu2017nature}. 

In contrast, for unstable water film, $F_{\text{s}} = 1.9$ $\mu$N is much larger than $F_{\text{k}} = 150$ nN. In this case, $F_{\text{s}} $ is the force required to force wet the water film (Supplementary Movie S2) and its magnitude ($\sim$ $\mu$N) is comparable to $F_{\text{fric}}$ in lotus-leaf effect surfaces. Once the droplets moves, the water film is dynamically stabilized at a thickness given by the LLD scaling (Eq.~\ref{eq:LLD}) and $F_{\text{k}}$ is once again given by Equation \ref{eq:F_LLD}. Experimentally, we found that $F_{\text{s}}$ is independent of the applied $U$ (solid line in Figure \ref{fig:friction}c), similar to the friction behaviour of a water droplet on superhydrophobic lotus-leaf effect surfaces \cite{daniel2018origins, liu2017nature}.   

The adhesion force $F_{\text{adh}}$ can similarly be measured using the DFA, with the surface positioned vertically (Fig.~\ref{fig:adh}a). In a typical experiment, the surface first approached the droplet from afar up to a loading force $F_{\text{load}}$ (A to B in Figure \ref{fig:adh}b), before retracting at a controlled speed $U$ (B to C). The adhesion force $F_{\text{adh}}$ is the maximum force just before the droplet detaches (Supplementary Movies S3 and S4). For a droplet of $2R =$ 2.0 mm and $U$ = 50 $\mu$m/s, $F_{\text{adh}}$ = 85 nN and 4.6 $\mu$N for stable and unstable water film, respectively (Fig.~\ref{fig:adh}b). For the latter case, we occasionally measured a snap-in force $F_{\text{snap}}$ as the water film dewetted and the oil droplet contacted the surface; there is contact line pinning and the droplet's base is jagged and distorted. After the droplet detached, a small microdroplet was sometimes left on the surface. In contrast, for stable water film, $F_{\text{snap}}$ was never observed, the droplet's base was always smooth and circular in shape, and there was no microdroplet left behind (See RICM images at time point C in Figure \ref{fig:adh}b).     

\begin{table*}[]
\caption{\label{tab:hys} Liquid-repellent properties of different surface classes \cite{daniel2017oleoplaning, daniel2018origins, keiser2017drop, reyssat2010dynamical, reyssat2009contact}. $\phi$ is the solid surface fraction.}. 
\begin{ruledtabular}
\begin{tabular}{cccccc}
	Surface class & $F_{\text{fric}}$ & $\lvert F_{\text{fric}} \lvert$ &$F_{\text{adh}}$ &$\lvert F_{\text{adh}} \lvert$ &Comments\\
	\hline
	Lotus-leaf effect &$\sim 2r \gamma \phi^{1/2}$ or  & 1--10 $\mu$N &$\sim 2r \gamma \phi^{1/2}$ or  & 1--10 $\mu$N &  no dependence on $U$ \\
	&$\sim 2r \gamma \phi \ln \phi$ & & $\sim 2r \gamma \phi \ln \phi$ & & low friction and low adhesion  \\ \\         
	Lubricated & $\sim 2r \gamma Ca^{2/3}$ & 0.1--5 $\mu$N &$\sim 2r \gamma$ & $>$ 10 $\mu$N & increases non-linearly with $U$ \\
	           & & & & & low friction but high adhesion 
	\\ \\ 
	Polyzwitterionic & $\sim 2r \gamma Ca^{2/3}$ & 0.02--1 $\mu$N & $ \sim 2R \gamma Ca^{0.6}$ & 0.02--1 $\mu$N & increases non-linearly with $U$\\
			   brush & & & & & very low friction and adhesion
	 
\end{tabular}
\end{ruledtabular}
\end{table*}

For a stable water film, $F_{\text{adh}}$ is once again dominated by viscous forces. At the point of droplet detachment, using scaling arguments, we expect $h_{\text{water}} \sim R \, Ca^{1/2}$ and $r \sim R \, Ca^{1/4}$ (cf. Equation \ref{eq:LLD}) \cite{bouwhuis2012maximal, klaseboer2014universal} (See Supplementary Figure S9 for derivation). Hence, $F_{\text{adh}}$ should scale as 
\begin{equation} \label{eq:F_adh}
\begin{split}
F_{\text{adh}} &= (2\gamma/R) \, \pi r^{2}\\
               &\sim 2R \gamma Ca^{1/2}.
\end{split}
\end{equation} 
However, we found that the experimental data is best explained with an exponent of 0.6 (rather than 0.5) and a prefactor of 4 (dashed line in Figure \ref{fig:adh}c). To account for this discrepancy requires solving the Navier-Stokes equation fully, but the physical rationale for Equation \ref{eq:F_adh} is nevertheless correct: $F_{\text{adh}}$ is viscous in origin and is therefore insensitive to the details of the polymer brush layer, e.g. its thickness $h_{\text{wet}}$ = 9--45 nm. Experimentally, we also found that $F_{\text{adh}}$ is independent of $F_{\text{load}}$ (Supplementary Fig.~S10). At the lowest velocity $U$ = 2 $\mu$m/s, we obtained $F_{\text{adh, min}}$ = 20 nN for a millimetric droplet. In comparison, when the water film is unstable, $F_{\text{adh}}$ is much larger ($> 1$ $\mu$N) and is independent of $U$ = 4--2000 $\mu$m/s and $Ca$ = $8 \times 10^{-8}$--$4 \times 10^{-5}$ (solid line in Figure \ref{fig:adh}c).  

Finally, we note that the friction and adhesion forces measured by the DFA are primarily due to contact-line pinning or viscous flow at the droplet's \textit{base}, and not due to flow around the droplet or any capillary force acting on the cantilever. We confirmed this by repeating the friction force measurement with the droplet's base far from the surface (a couple of millimetres away), in which case no measurable cantilever deflection was detected. See also discussion in Supplementary Figures S6 and S7.  

\section*{Discussion}

We would like to point out that the polyzwitterionic brush surfaces are \textit{qualitatively} different from other classes of liquid-repellent surfaces, notably the lotus-leaf effect surfaces \cite{quere2008wetting, reyssat2010dynamical} and the \textit{Nepenthes} pitcher-plant inspired lubricated surfaces \cite{lafuma2011slippery, wong2011bioinspired}. For example, for polyzwitterionic brush surfaces $F_{\text{fric, adh}}$ varies non-linearly with $U$ ($\propto U^{2/3}$ and $U^{0.6}$, respectively), whereas for lotus-leaf effect surfaces $F_{\text{fric, adh}}$ is relatively insensitive to $U \sim $ mm/s or less \cite{reyssat2010dynamical, liimatainen2017mapping, daniel2018origins, mouterde2019superhydrophobic}. Table \ref{tab:hys} summarizes the functional forms of $F_{\text{fric, adh}}$ for three surface classes, as reported here and elsewhere in the literature \cite{daniel2017oleoplaning, daniel2018origins, keiser2017drop, reyssat2010dynamical, reyssat2009contact}. The values of $\lvert F_{\text{fric, adh}} \lvert$ are for droplets of size 2$R$ = 1--4 mm and speeds $U$ = 0.01--2 mm/s, and as discussed previously, $\lvert F_{\text{fric, adh}} \lvert$ is much lower for polyzwitterionic brushes as compared to lotus-leaf effect and lubricated surfaces.   

In this work (and our previous work \cite{daniel2018origins}), we have presented the friction and adhesion force data in their non-dimensionalized forms: $F_{\text{fric}}/2r \gamma$ or $F_{\text{adh}}/2R \gamma$ vs. $Ca$. Figures \ref{fig:friction}c and \ref{fig:adh}c can be thought of as universal phase diagrams for liquid-repellency; $F_{\text{fric}}$ and $F_{\text{adh}}$ data for different classes of liquid-repellent surfaces and experimental conditions (different oils, droplet sizes, and speeds) can be plotted on the same graph for direct comparison. The non-dimensional quantity $F_{\text{fric}}/2r \gamma$ is also numerically equivalent to the contact angle hystereses $\Delta \cos \theta$ that are typically reported in the literature (Furmidge's relation) \cite{furmidge1962}. This, again, allows for easy comparison with literature data. For example, most lotus-effect surfaces have $\Delta \cos \theta$ = 0.05--0.10, much higher than the values reported here where $F_{\text{fric}}/2r \gamma$ = 0.001--0.01 \cite{daniel2018origins}.  

\section*{Conclusion}
In summary, we have clarified the origin of underwater oil-repellency for polyzwitterionic brush surfaces, in particular the role played by hydration lubrication. We were able to measure the thickness of the water film beneath the oil droplet with nanometric accuracy, as well as the friction and adhesion forces with nN resolutions using a custom-built Droplet Force Apparatus (DFA). Just as the Surface Force Apparatus has revolutionized the study of surface forces, we believe our DFA setup will help resolve many of the outstanding questions in wetting science. Finally, while we have confined our discussion to oil-repellent properties of polyzwitterionic brushes, many of the ideas and techniques outlined here are relevant to other oil-repellent surfaces (e.g polycationic/anionic brushes and hydrogels), and more generally to other classes of liquid-repellent surfaces. The insights generated here are pertinent to the rational design of anti-fouling materials not just for liquids, but also for biological contaminants.   

\section*{Acknowledgements}

We thank R.E. Cohen and P.V. Braun for fruitful discussions, and K. Serrano for advice on polymer brush synthesis. The authors are grateful to the Agency for Science, Technology and Research (A*STAR) for providing financial support under the PHAROS Advanced Surfaces Programme (grant number 1523700101, project number IMRE/16-2P1203).  

\section*{Authors' Contributions}

D.D. and N.T. conceived and planned the experiments. D.D., A.Y.T.C., L.C.H.M., R.L., X.Q.K., and X.Z. executed the experimental work. D.D. analysed the experimental results and performed the scaling analysis for the friction and adhesion forces. D.D. and N.T. wrote the manuscript. All authors reviewed the manuscript.

\newpage
\providecommand{\noopsort}[1]{}\providecommand{\singleletter}[1]{#1}%

\clearpage
\section*{Materials and Methods}

\textbf{Materials} 
\textit{Chemicals for brush growth:} (3-trimethoxysilyl)-propyl 2-bromo-2-methylpropionate (Br-initiator, 95 \%) was purchased from Gelest Inc. [2-(Methacryloyloxy)ethyl]dimethyl-(3-sulfopropyl) ammonium hydroxide (SBMA, $\geq$ 96 \%), copper(I) bromide (CuBr, $\geq$ 98\%), and 2,2\'-bipyridine (bipy, $\geq$ 99 \%) were purchased from Sigma-Aldrich. Methanol was purchased from J.T. Baker. All chemicals were used as received. 

\textit{Oils:} Silicone oil (Polyphenyl-methylsiloxane, viscosity $\sim$ 100 mPa.s, density 1.06g/ml) and fluorinated oil (viscosity $\sim$ 15 mPa.s, density 1.93 g/mL, Fluorinert FC-70) were purchased from Sigma-Aldrich. The water-oil interfacial tensions are 40 and 50 mN/m for silicone and fluorinated oils, respectively, as measured using the pendant drop method. 

\textit{Water:} Deionized water with a resistivity of 18.2 M$\Omega$.cm was obtained from a Milli-Q water purification system (Millipore, Bedford, MA, USA). \\
  
\textbf{Sample preparation.} The polymer brush surfaces are prepared using surface initiated Atom Transfer Radical Polymerization (ATRP) using a protocol adapted from \citeauthor{azzaroni2006ucst} (2006) \cite{azzaroni2006ucst}.

\textit{Initiator monolayer deposition:} The surfaces (glass or silicon wafer) were rinsed extensively with deionized (DI) water, and then ethanol, before drying under a nitrogen stream. The dried surfaces were then subjected to oxygen plasma surface cleaning at 150 W for 120 s. Br-initiator was vapour deposited onto the cleaned surfaces. In a typical procedure, the cleaned surfaces were heated in a vacuum oven at 75$^{\circ}$C with the Br-initiator (100 $\mu$L) overnight. The silanized surfaces were then cleaned (by rinsing with anhydrous toluene, ethanol, and water, sequentially) and then dried under a nitrogen stream. The dried silanized surfaces were then heated in an oven at 110$^{\circ}$C for 20 minutes.   

\textit{Polymer brush growth:} In a typical procedure, the solvent solution (4:1 methanol:water, 50 mL) was first deoxygenated by bubbling with nitrogen for at least 30 minutes. SBMA (53.70 mmol, 15.0 g) was dissolved in 40 mL of the solvent solution to form the monomer solution, while bipy (1.344 mmol, 209.88 mg) and CuBr (0.5375 mmol, 77.11 mg) were dissolved in the remaining 10 mL of the solvent solution to form the catalytic solution. Both solutions were then stirred, while continuously being bubbled with nitrogen. After about 10 minutes, the catalytic solution was added to the monomer solution and was allowed to stir for another 2 minutes under nitrogen protection. The reaction mixture was then transferred to the reaction vessel containing the silanized surfaces. The reaction was performed under nitrogen protection. The reaction time was varied to achieve various brush heights. Upon completion of the polymer brush growth, the surfaces were rinsed with copious amounts of warm DI water (60$^{\circ}$C) and dried under a nitrogen stream.    

\textit{Chemical characterization} was performed using X-ray photoelectron spectroscopy (XPS, Thermo Scientific Theta Probe system) to confirm the expected chemical structure (See Supplementary Figure S2)\\  

\textbf{Determining brush thickness.} The dry and wet thicknesses of the polymer brush layer on a silicon wafer can be determined using either ellipsometry or Atomic Force Microscopy (AFM) measurements. Both techniques give similar results. For brush layer on glass, the thickness can only determined using AFM measurement, as the refractive index contrast between the brush layer and glass is too small for accurate ellipsometry measurement.     

\textit{Ellipsometry measurement:} Spectroscopic ellipsometry measurement was performed using a commercial ellipsometer (VASE from J.A. Woollam Co, Inc) at an incident angle of 70$^{\circ}$. For ellipsometry measurement in water, the surface is placed in a custom built quartz chamber with side windows at 70$^{\circ}$ angle. The index of refraction of the polymer brush layer was fitted using a Cauchy model (dry and wet). For swollen brush layer, the index of refraction can alternatively be modelled using the Bruggeman effective medium approximation. Both models (Cauchy and Brugemann) give similar results. At least three measurements across each sample were taken.

\textit{AFM measurement:} A small section of the brush layer was first scratched using a tweezer to reveal the underlying glass/silicon substrate; the surface was then rinsed with copious amount of water and dried under nitrogen gas to remove any detached brushes. The dry and wet thicknesses of the brush layer can then be measured using AFM by probing the scratched section. Detailed topography of the brushes was obtained using the Quantitative Imaging mode (QI-Mode, JPK Instruments), which minimizes the distortion due to lateral and compressive forces \citep{rodriguez2015quantifying, chopinet2013imaging}. This is particularly important for brushes under water, which become soft when swollen (See Supplementary Figure S3 for details). \\

\textbf{Determining water film thickness}. The thickness of the water film can be determined using either microscopic reflection spectroscopy or dual-wavelength confocal Reflection Interference Contrast Microscopy (RICM). See Supplementary Figures S4, S5, and S8.  

\textit{Confocal RICM:} Briefly, we raster scanned the surface simultaneously with two focused beams of monochromatic lights with wavelengths $\lambda = 458$ and 561 nm, and captured the reflected light through the pinhole of a confocal microscope; as a result, only reflected light from the focal plane, i.e. the interface of interest, was able to reach the photomultiplier tube of the microscope. This is crucial, because the weak refractive index contrast leads to a weak reflection signal that can be overwhelmed by stray light. In the presence of a thin water film, the light reflected off the various interfaces (glass-brush, brush-water, and water-oil) will interfere with one another constructively or destructively and result in bright or dark fringes, respectively. From the reflection intensities of the two wavelengths, the water film profile $h_{\text{water}}$ can be deduced unambiguously using classical Fresnel theory \cite{born2013principles}.  

\textit{Microscopic reflection spectroscopy:} In this method, white light is used instead of monochromatic lights. The light reflected off the various interfaces will again interfere with one another and its intensities (at various wavelengths) were then analyzed using a spectrometer. As with confocal RICM, the local film thickness $h_{\text{water}}$ can deduced using classical Fresnel theory. See Supplementary Figure S4 for the optical path of the setup.

\textit{Fitting with Fresnel theory:} In both methods, the reflection intensities were fitted to classical Fresnel theory using Levenberg-Marquart algorithm with $h_{\text{water}}$ as the only fitting parameter.   

The refractive indices of the various materials (water, polymer brush, glass, and oil), including their optical dispersions, as well as the the polymer brush thickness $h_{\text{wet}}$ were measured using an ellipsometer and inputted directly into the Fresnel theory. The accuracy of $h_{\text{water}}$ is therefore limited by the uncertainty of $\Delta h_{\text{wet}}$, which is about 5 nm, much smaller than $h_{\text{water}} >$ 100 nm.        

\section*{Data availability}
The authors declare that the data supporting the findings of this study are available from the corresponding author upon reasonable request.

\end{document}